# AUTONOMOUS UNDERGROUND FREIGHT TRANSPORT SYSTEMS – THE FUTURE OF URBAN LOGISTICS?

L. BIENZEISLER [a], T. LELKE [a] and B. FRIEDRICH [a]

[a] *Institute of Transportation and Urban Engineering, TU Braunschweig, Hermann-Blenk-Straße 42, 38108 Braunschweig, Germany*
*Email: l.bienzeisler@tu-braunschweig*

## ABSTRACT

We design a concept for an autonomous underground freight transport system for Hanover, Germany. To evaluate the resulting system changes in overall traffic flows from an environmental perspective, we carried out an agent-based traffic simulation with MATSim. Our simulations indicate comparatively low impacts on network-wide traffic volumes. Local $CO_2$ emissions, on the other hand, could be reduced by up to 32 %. In total, the shuttle system can replace more than 18 % of the vehicles in use with conventional combustion engines. Thus, an autonomous underground freight transportation system can contribute to environmentally friendly and economical transportation of urban goods on the condition of cooperative use of the system.

## 1.    INTRODUCTION AND RELATED WORK

Despite the rapidly increasing number of parcels and the induced delivery trips, especially in urban areas, the modeling of urban freight traffic has played a minor role in traffic modeling over the past decades. However, the topic recently started to receive more sophisticated attention (Reoffer et al., 2018). Fueling this development, the emergence of agent-based transport models, such as the framework MATSim (Horni et al., 2016), has enabled more detailed modeling and evaluation of urban freight traffic. A substantial part of commercial traffic is the continuously growing courier, express, and parcel (CEP) segment (Schroeder et al., 2015), along with an increase in e-commerce. Due to the consequential increase in delivery traffic, new and innovative approaches or concepts for urban logistics are needed to avoid the congestion effects of urban infrastructures.

To reduce overall traffic volume, delivery traffic could be transferred to an automated underground freight transport system (*AUFT*). Parcels and freight would be transported from the production site to the logistics hub at the city borders or directly to the stores in the city center. The system provides an opportunity to address both the impact of freight traffic disruptions on urban traffic and freight traffic disruptions caused by road traffic. Therefore, the entire supply chain becomes more plannable, reliable, and flexible. With *AUFT,* a high degree of schedule reliability might be achieved. In addition, operating in a tunnel makes the system space-saving and weather-independent (OECD, 2003). Apart from these advantages over conventional logistics systems, however, underground freight transport systems require significantly higher investment costs. They also need a complex operational planning system coordinated with conventional logistics systems.

Various concept studies can be found in literature (Rijsenbrij et. al, 2006, Cui & Nelson 2019, Shahooei et al., 2019, Hai et. al 2020, Hu et. al 2022). However, the detailed effects of these systems on urban traffic flow have not been studied. We thus analyze the impact of implementing a freight shuttle on overall traffic flow as indicated by daily trip distances and mileage of delivery traffic in a simulation case study. As a secondary outcome, we assessed the impact on $CO_2$ emissions.

## 2.    STATE OF THE ART

The first approaches of autonomous transport vehicles (so-called Automated Guided Vehicles) were applied in the USA in the early 1950s. There are various technical approaches and solutions to realize such an *AUFT* concept. An advanced project for a new logistics system is Cargo Sous Terrain (CST) from Switzerland[1]. The system is characterized by its modular and intermodal structure, whose core element is *AUFT*. Along with the German-developed systems CargoCap[2] developed as a rail-based system and Smart City Loop[3] with a technically analogous concept to CST, the HyperLoop[4] concept receives a high degree of global attention. The system differs from all other systems by operating at significantly higher speeds (over 1,000 km/h) in tubes with reduced air resistance using magnetic levitation technology.

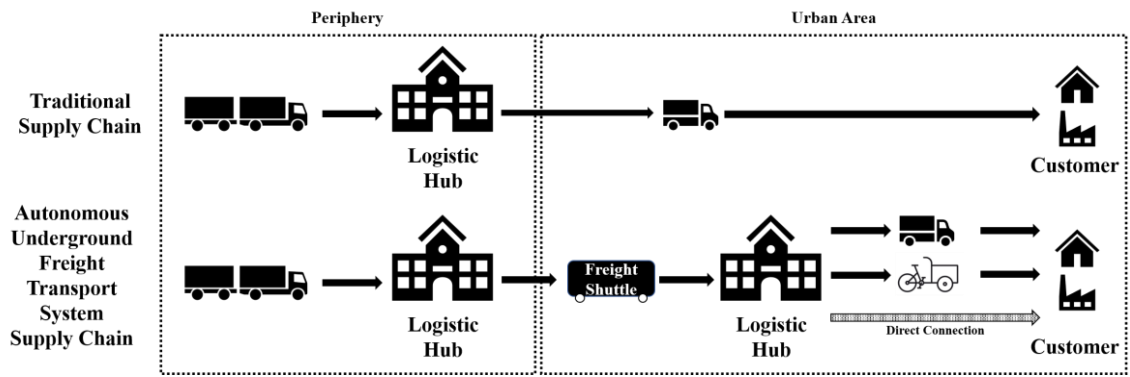

Figure 1. Process in comparison between a traditional supply chain and a supply chain of an autonomous freight transport system.

The *AUFT* is independent due to its complete guidance in the tunnel system, separating parts of the freight transport from the surrounding traffic. Electric and autonomous operated vehicles are used in tunnels capable of carrying various goods. Outside the city center, the cargo is transferred from trucks to the underground system at hubs in strategically convenient locations. Already existing logistics infrastructure (roads etc.) is used. City hubs are created as close as possible to the end customer. The vehicles from the tunnel reach these goods handling points automated via ramps or elevators. The distribution of goods in the inner city is directly carried out from there by cargo bike or conventional delivery vehicles. In addition, direct connections of companies can avoid the transfer of goods within the city. The fundamental principle and associated supply chains are shown in Figure 1.

The work of Lampert (2019) provides a framework for our work and for our implementation of the simulation concept. We first applied the developed concept for an *AUFT* for Hanover, Germany. We then successively expanded and optimized the simulation and developed different scenarios. The new system can potentially replace truck transports of the penultimate mile. It connects the logistics centers outside the city with the city hubs in the city center via underground pipelines. As a result, a higher level of performance and safety is expected.

## 3.    CONCEPT

We selected the technical concept based on the existing approaches that resulted in different simulation scenarios. The Hyper Loop concept, with its high system speed, was unsuitable due to the short system distances in Hanover impeding efficient use. With developmental advantages, and a more flexible and expandable system, the Hanover *AUFT* originates from the technical framework of the CST project already deployed in Switzerland. The *AUFT* uses small autonomous vehicles which can carry two European standard pallets (1.2 x .8 m). Approximately 80 parcels can be accommodated on one pallet. Since the pallet cannot always be optimally utilized due to bulky goods, 70 packages per unit were assumed, resulting in a total capacity of 140 parcels per vehicle. Since no or few personnel costs are incurred, significantly lower costs per time and distance can be expected.

We decided to use a demand-based methodology to develop the mode of operation and route planning. Three industry sectors were selected as potential customers for the system: supply and delivery traffic

---

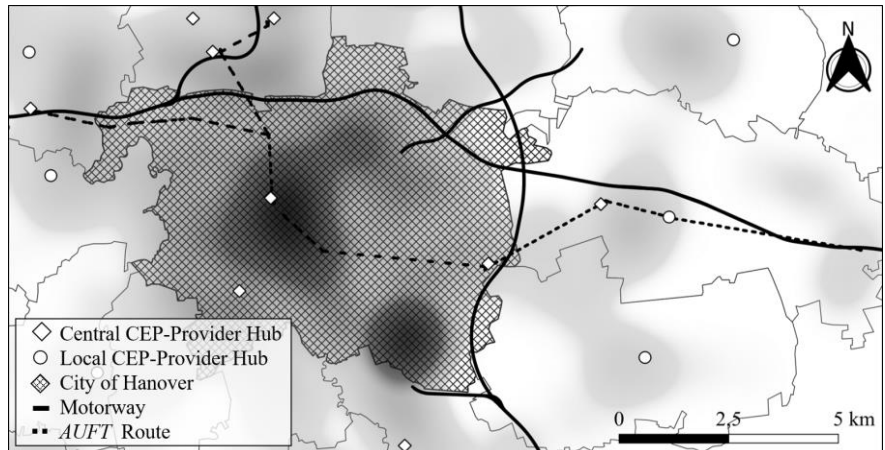

Figure 2. Planned Route of the *AUFT* System. The heat map shows the extrapolated parcel volume for the Hanover region. The darker the representation, the higher the local parcel volumes.

of the CEP sector, supply traffic of the retail sector, and supply traffic of automotive companies. Thus, an analysis of the existing logistics infrastructure, commercial and retail centers, as well as population structure was carried out to determine the optimal routing of the *AUFT*.

As detailed freight transport demand data are not publicly available, different available data sets and derived statistical distributions were used to estimate demand and the resulting freight trips in the Hanover region. The demand for CEP or the volume of parcels per road element was extrapolated based on a real-world data set for different service providers operating in the Hanover region. Deliveries to parcel lockers and misdeliveries returning to parcel stores were also integrated into our CEP demand. For the industry sector, we reduced the complexity of the estimation to the largest automotive companies receiving a direct connection to the *AUFT*. While the vehicles produced at the car manufacturer are transported mainly by rail, most goods are supplied by truck. Using available data crawled from company websites as well as factory areas and employee numbers, industry supply chains and thus resulting freight trips were estimated.

The data were used as input for route planning. Hub locations outside the city were located near already existing distribution centers using their infrastructure to reduce investment costs as well as transport distances. In addition, five hubs or distribution centers, so-called micro depots, were integrated into the shuttle route allowing goods to be transshipped and stored. From these depots, direct delivery to the surrounding area either with the conventional delivery vehicle or with cargo bikes is possible (Llorca & Moeckel, 2021). In addition, industry companies and a shopping center in the city center are directly connected to the system. These specifications resulted in a planned route (see Figure 2) which covers a distance of 44 km in total. Consequently, the system operates in areas of the city where high parcel volumes are expected.

## 4. IMPLEMENTATION

The developed *AUFT* concept allows the logistic service providers to operate micro depots within the city area. AUFT enables parcel deliveries to be distributed directly to the surrounding area without additional transshipment processes. The structure provides the capability to restructure the supply chains: Goods and parcels can be shipped directly from the shuttle depot to the end customer without being distributed via external warehouses. In total, 5 out of 7 of the logistic service providers included in our model are located in the feeding area of the *AUFT* and thus have access to it. We assumed a hub capacity of 4,000 parcels per hub and per day (one truckload per logistic supplier with a delivery volume of 800 parcels).

This planning allowed us to implement three different distribution use cases for the last-mile delivery to compare the results with a basecase scenario (BC) representing the status quo of Hanover's logistic system:

### Scenario 1: Separated Hub Usage (SHU)

Each logistics service provider was assigned with a storage capacity of 800 parcels per hub of the freight shuttle for direct delivery to the surrounding end customers using conventional delivery vehicles. The freight shuttle operator ensured that the parcels arrived at the corresponding hub by the beginning of the tour.

**Scenario 2: White-Label Hub Usage (WHU)**

All delivery services adjacent to the *AUFT* were each assigned to the closest micro hub of the freight shuttle until the capacity limit of 4,000 parcels per hub was reached. Deliveries were not differentiated by the provider anymore to enable optimized tour planning. The role of the freight shuttle operator changed from a supply to a delivery agent.

**Scenario 3: White-Label Hub Usage with Cargo Bikes (WHU-B)**

This scenario represents an extension of the white label scenario. Parcels could also be delivered by cargo bike in the vicinity of the hubs. The freight shuttle operator can decide whether a bike or a conventional delivery vehicle is used.

To investigate the effects of the simulation scenario, we carried out an agent-based traffic simulation based on an existing model of Hanover (Bienzeisler et al., 2019). Using the agent-based traffic simulation tool MATSim in combination with the freight contribution by Schröder et al. (2012), we included the planned *AUFT*. Then, we compared it to a baseline scenario with comparable parameters but without *AUFT* implementation.

We developed a two-stage simulation concept schematically diagrammed in Figure 3. As an initial input and as a basecase scenario, we used the freight demand consisting of parcel deliveries. It includes hub supply deliveries and industry supply trips. Each company involved in the demand structure was modeled as a so-called carrier agent to whom its related services were assigned. For example, all parcels in a delivery area were connected to the associated logistics supplier. Subsequently, the vehicle routing problem (VRP) was solved for each carrier agent with the external java library *jsprit*[5] using free flow travel times and vehicle information (see table 1). This vehicle information is based on Zhang et al. (2018). As a result, we obtained scheduled tours and the number of vehicles used.

These generated carrier plans were subsequently simulated with MATSim to include the impact of individual traffic in the model. To avoid excessive computation times for the simulation model, the freight traffic was not simulated with the individual traffic. Instead, the influence of individual traffic was reproduced by so-called network change events. Hourly average link speeds obtained from the MATSim individual traffic model were used and assigned to the freight network to emulate network-wide link loads. We generated plans for each last mile delivery carrier representing the input for the supply of each hub with the freight shuttle.

Since we have not implemented the transfer of goods between the systems directly in our model, the simulation of the freight shuttle carrier had to be performed independently and after the simulation of the last mile carriers. The carrier plan of the freight shuttle carrier depends on the optimized and previously simulated tours of the last mile carriers. Temporal and spatial departures of the last-mile vehicles provided the delivery window for the freight shuttle. All cargo must have been delivered by the shuttle to the corresponding hub just in time to avoid delays in the operational process. As an actual use case, this process corresponds to a tour schedule provided at the beginning of the day by each last-mile carrier as a demand to the freight shuttle operator. This operator ensures that the cargo is transported on time to the hub or, depending on the scenario, directly to the customer. In addition, industrial and retail jobs were handled using the freight shuttle. Since all modeled industrial facilities are directly connected to the freight shuttle, no additional deliveries were necessary.

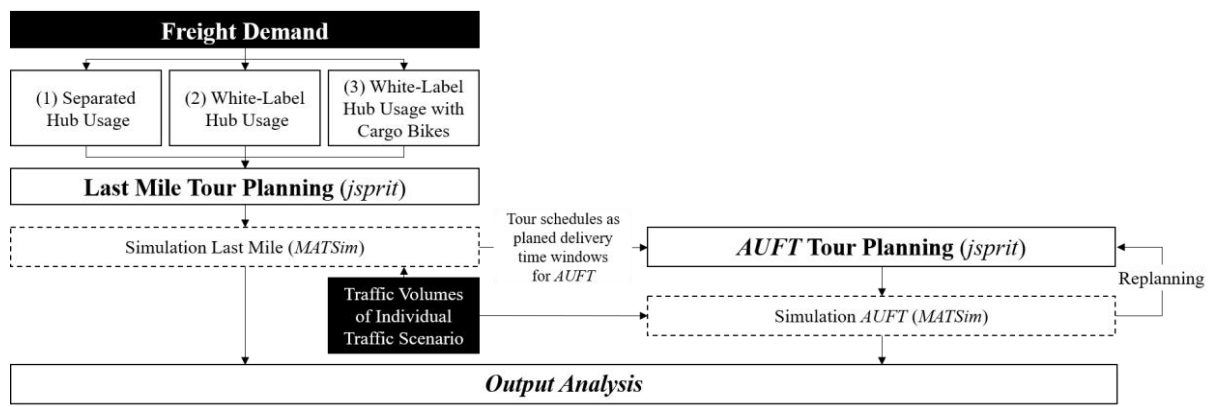

Figure 3. Flowchart of the *AUFT* implementation in MATSim based on the concept of Lampert (2019).

The assigned deliveries of the freight shuttle were modeled as shipments with time windows in *jsprit,* enabling an intelligent connection of jobs at different hubs to optimize the freight shuttle operation. Last-mile vehicles could not depart until their cargo had arrived at their hub. To avoid a delay in the operational process and to minimize the storage time, we implemented two requirements:

(1)  At the earliest one hour before departure, the cargo should arrive at the hub
(2)  At least 15min before departure, all cargo of the scheduled tour should have arrived at the hub

Failure to comply with this time window was heavily sanctioned by penalties for the freight shuttle carrier. Finally, the *AUFT* traffic assignment was done with MATSim in a second simulation. To keep the computation times of VRP with *jsprit* within a feasible time range, we divided the number of carriers operating the *AUFT*. The assigned shipments were distributed evenly among them. Thus, an optimum solution for the VRP is not ensured, but this simplification is acceptable due to the high number of orders per carrier. Relying on previous research (Bienzeisler et al., 2021), we carried out a total of three simulation runs per scenario and determined the mean values of the characteristic simulation data.

Table 1. Vehicle costs and capacities.

| Vehicle Type | Costs (€/m) | Costs (€/s) | Fixed Costs (€) | Capacity |
|---|---|---|---|---|
| CEP-Vehicle | 0.00037 | 0.0063 | 48.8 | 230.0 |
| CEP-Cargo-Bike | 0.000103 | 0.0033 | 3.27 | 23.0 |
| Supply-Truck | 0.00086 | 0.008 | 140.0 | 800.0 |
| Freight Shuttle | 0.00035 | 0.002 | 30.0 | 140.0 |

## 5.    SIMULATION RESULTS

To evaluate the impact of the *AUFT* implementation within the different scenarios, we analyzed the resulting last-mile tours as the primary outcome. Figure 4 shows the distribution of the utilization of all vehicles operating above ground within the supply chain in relation to the corresponding tour length. The simulation of the basecase indicates that the conventional delivery strategy achieves a comparatively high utilization of the vehicles. However, the analysis also shows a share of underutilized vehicles, especially in the distance range of typical CEP tours (approx. 30 to 70 kilometers). This share is highest in scenario 1 (Separated Hub Usage), particularly for shorter tours. The reason for this is the non-cooperative use of *AUFT's* hubs. As a result of the comparatively smaller number of assigned parcels, the tours cannot be optimally bundled. There is a larger share of vehicles that are not fully utilized. In the current modeling state, the algorithm cannot pass these shipments back, resulting in non-efficient tours to the actual logistic service provider. The introduction of the white label concept in scenarios 2 and 3 then demonstrated a higher utilization through the aggregation of trips at *AUFT's* hubs. The introduction of delivery by cargo bicycles for parts of the last mile resulted in a reduction of the tour length. The cargo bikes cover a shorter distance and have a lower load capacity. As a result, they can always start their tour at total capacity.

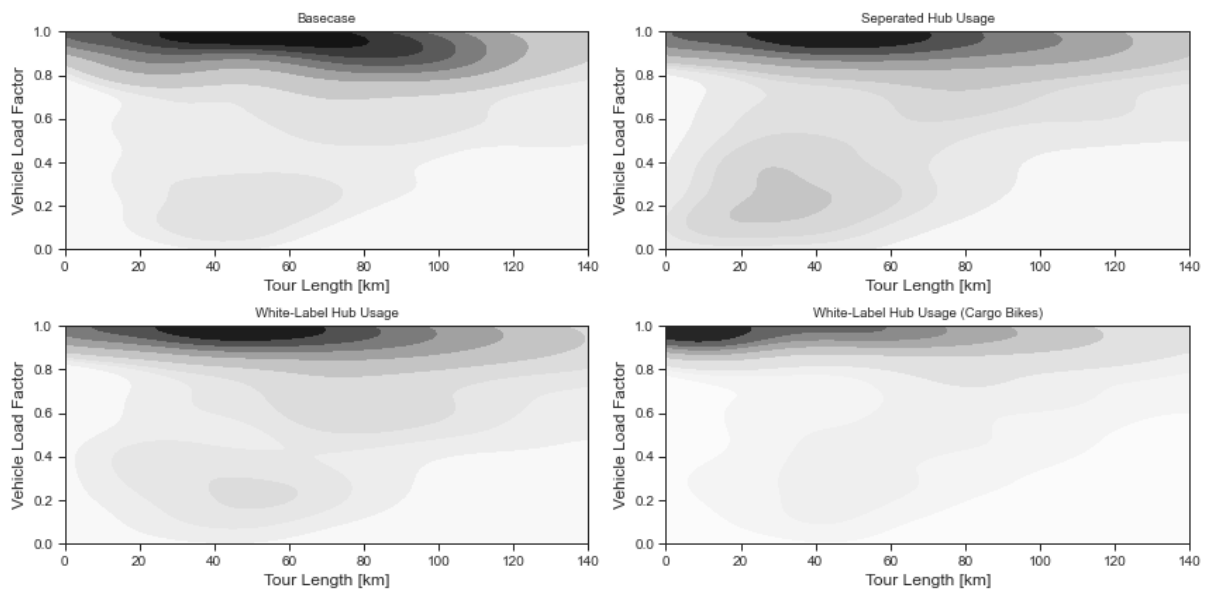

Figure 4. Kernel density estimation of tour length and associated vehicle loading factor distribution for each scenario. The darker the representation, the higher the number observations in our dataset.

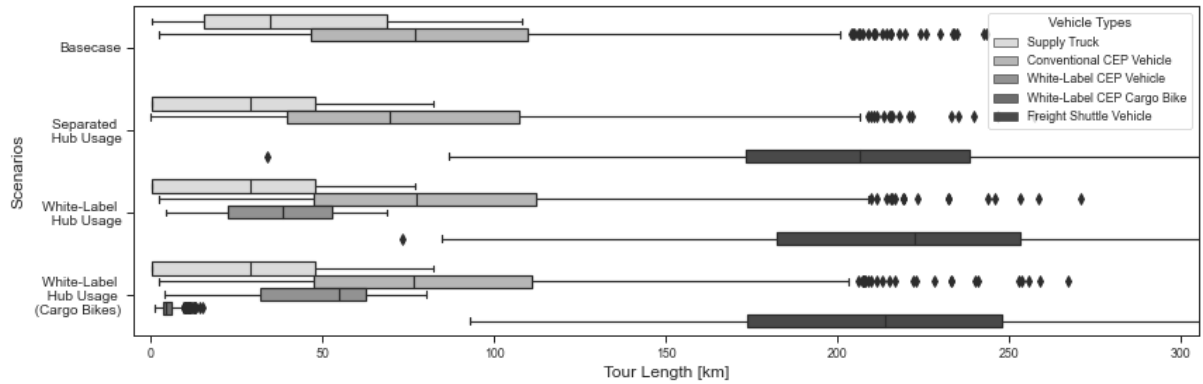

Figure 5. Distribution of daily tour lengths differentiated by vehicle type.

When considering the tours introducing the *AUFT* in detail (see Figure 5), the supply tours of the hubs become shorter compared to the basecase scenario. Supply goods can be shipped via the shuttle system. Freight shuttle vehicles, on the other hand, cover longer daily distances of over 200 km. Incorporating new hubs into the supply chain shortens conventional parcel delivery routes alike. Due to the more homogeneous distribution of shipments, the tours can be further optimized in the white-label case. The supply tours of the hubs change and become shorter, as most of the supply goods are shipped via the shuttle system. On the highways and major roads leading into the city center, this development leads to a reduction in traffic congestion in all three scenarios. As an example, Figure 6 shows the changes in network loads for scenario 3. More than 125 truck trips can be saved every day on the highway in the north of the city, connecting the warehouses of the logistics service providers with the city center. An increase in traffic volumes can be observed directly at the hubs as the delivery vehicles depart from here. In the case of scenario 3, this increase is particularly significant. We can attribute this fact to the high number of cargo bikes traveling on these sections of the network. However, since the bikes operate on a separate bike network, they have no further impact on other vehicles.

According to our results, around 22,000 km of vehicle mileage and 400 driven truck hours could be saved with the introduction of the new freight system per day. To estimate the environmentally relevant impacts as a secondary outcome, we converted the covered vehicle kilometers into $CO_2$ emissions. To calculate these emissions, the values 197.295 g per km for light commercial vehicles and 789.505 g per

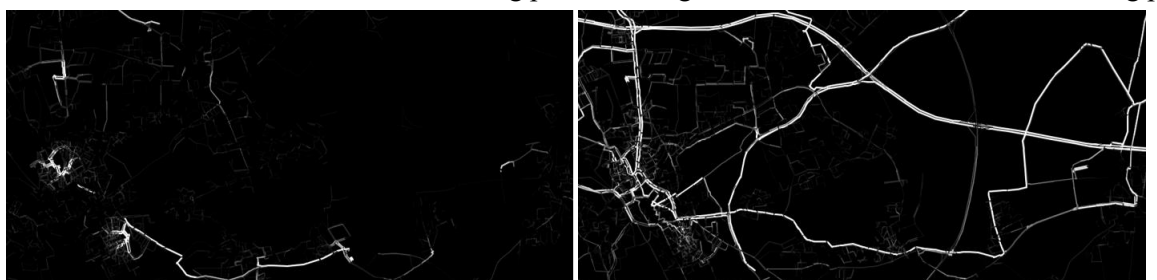

**Additional Traffic in the Network**          **Reduced Traffic in the Network**

Figure 6. Additional and reduced network loads in comparison between basecase and scenario 3 (White-Label Hub Usage with Cargo Bikes).

km for heavy-duty vehicles were used (Notter et al., 2020). The number of used vehicles and the distances traveled differentiated by vehicle class and scenario is shown in Figure 7. In total, the introduction of an *AUFT* leads to a reduction in $CO_2$ emissions of 30.7 % in scenario 1 (SHU), 31.1 % in scenario 2 (WHU), and 31.7 % in scenario 3 (WHU-B) (see Table 2). Our data also show that the *AUFT* vehicles cover longer distances in the network than conventional supply trucks. Due to the significantly smaller loading size (17.5 % of a truck), the shuttle vehicles operate more frequently to deliver cargo. Yet these vehicles produce no local emissions as they are equipped with electric drives.

Table 2. Simulation results.

| Scenario | Total Distance [km] | Total Ground Distance [km] | Total Shuttle Distance [km] | Average Ground Vehicle Load [%] | Total Emissions CEP [t] | Total CO2 Emissions Trucks [t] | Total CO2 Emissions [t] |
|---|---|---|---|---|---|---|---|
| BC | 146,730 | 146,730 | 0 | 0.82 | 29.75 | 25.49 | 55.24 |
| 1. SHU | 214,430 | 124,343 | 90,087 | 0.78 | 29.40 | 8.88 | 38.28 |
| 2. WHU | 217,073 | 123,457 | 93,614 | 0.81 | 29.17 | 8.88 | 38.05 |
| 3. WHU-B | 220,484 | 122,179 2,556 (Bike) | 95,749 | 0.84 | 28.83 | 8.88 | 37.71 |

## 6. CONCLUSION

From a traffic and operational planning perspective, as well as from an environmental point of view, the simulations showed the advantages of an independent logistics infrastructure with an *AUFT* compared to mixed traffic with passenger traffic on the existing road infrastructure. As a result, traffic volumes could be reduced, especially on the feeder roads to the city. However, the results also show that the impact on overall traffic in the network is marginal. Thus, considering all saved fright trips, the freight traffic volume on selected roads in the city of Hanover can be reduced by $1 - 5$ %. Since loading and unloading operations are not included in our model, it can be assumed that the actual impact of the trips saved is higher than estimated by our model. It is uncertain if this influence could significantly increase the fundamental quality of traffic.

Moreover, the creation of new underground infrastructure requires significant investments. Our preliminary cost estimates have indicated that the large number of vehicles used in the *AUFT* will not reduce running costs either. We have calculated increasing running costs of $0.5 - 1.2$ % depending on the scenario. Using hubs with relatively small capacities in the city center in our developed concept, minor positive effects on traffic flow and volumes could be shown. In most cases, these resulted from the replacement of supply trips by the *AUFT*. However, economic studies must establish whether these positive effects will exceed the higher costs. Nevertheless, the *AUFT* offers particular advantages in the cooperation of different players in terms of the environment and life in the city. The efficiency of the system through cooperative use increases with the number of participating companies shipping their goods with the *AUFT*.

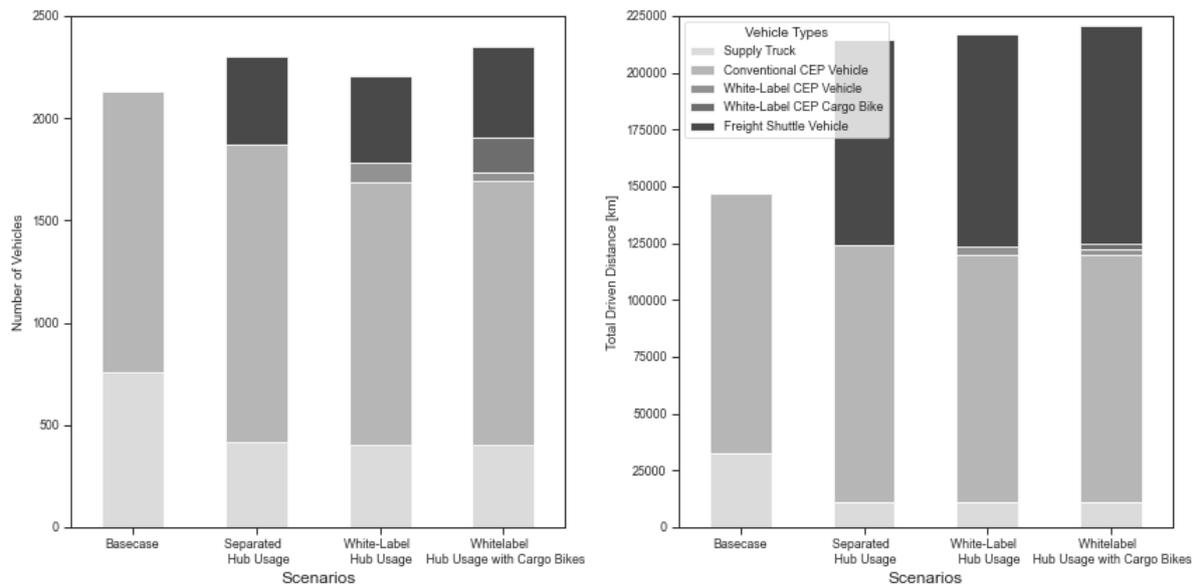

Figure 7. Number of vehicles used and distances by vehicle class differentiated by scenario.

For our future research work, we continue to include the total architecture and processes of the *AUFT* in our model. Currently, there is no physical transfer of goods at the shuttle hubs. However, by sanctioning the logistics service providers, the algorithm tries to optimize the tours to ensure punctual arrival. If this is not the case, the vehicles could still deliver the cargo to the customer.

**Acknowledgment**

The scientific research published in this article is granted by the Federal Ministry of Education and Research Germany for projects USEfUL XT (grant ID 03SF0609). The authors cordially thank the partners and funding agency.